# DAME: A Distributed Data Mining & Exploration Framework within the Virtual Observatory

Massimo Brescia[a*], Stefano Cavuoti[b] Raffaele D'Abrusco[c], Omar Laurino[d], Giuseppe Longo[b]

[a]*INAF – Osservatorio Astronomico di Capodimonte, Via Moiariello 16, 80131 Napoli, Italy*

[b]*Dipartimento di Fisica, Università degli Studi Federico II, Via Cintia 26, 80125 Napoli, Italy*

[c]*Center for Astrophysics - Smithsonian Astrophysical Observatory, 60 Garden Street, 02138 Cambridge Massachusetts USA*

[d]*INAF – Osservatorio Astronomico di Trieste, Via Tiepolo 11, 34143 Trieste, Italy*

**ABSTRACT**

Nowadays, many scientific areas share the same broad requirements of being able to deal with massive and distributed datasets while, when possible, being integrated with services and applications. In order to solve the growing gap between the incremental generation of data and our understanding of it, it is required to know how to access, retrieve, analyze, mine and integrate data from disparate sources. One of the fundamental aspects of any new generation of data mining software tool or package which really wants to become a service for the community is the possibility to use it within complex workflows which each user can fine tune in order to match the specific demands of his scientific goal. These workflows need often to access different resources (data, providers, computing facilities and packages) and require a strict interoperability on (at least) the client side. The project DAME (DAta Mining & Exploration) arises from these requirements by providing a distributed WEB-based data mining infrastructure specialized on Massive Data Sets exploration with Soft Computing methods. Originally designed to deal with astrophysical use cases, where first scientific application examples have demonstrated its effectiveness, the DAME Suite results as a multi-disciplinary platform-independent tool perfectly compliant with modern KDD (Knowledge Discovery in Databases) requirements and Information & Communication Technology trends.

[*] Corresponding author. Tel.: +39-081-5575-553; fax: +39-081-456710; e-mail: brescia@na.astro.it.

# 1. INTRODUCTION

Modern Technology in ICT (Information & Communication Technology) allows to capture and store huge quantities of data. Finding and summarizing the trends, patterns and outliers in these data sets is one of the big challenges of the information age. There has been important progress in data mining and machine learning in the last decade. Machine learning, data mining, or more generally KDD (Knowledge Discovery in Databases) discipline is a burgeoning new technology for mining knowledge from data, a methodology that a lot of heterogeneous communities are starting to take seriously. Strictly speaking KDD is about algorithms for inferring knowledge from data and ways of validating it. So far, the main challenge is applications. Wherever there is data, information can be gleaned from it. Whenever there is too much data or, more generally, a representation in more than 3 dimensions (limit to infer it by human brain), the mechanism of learning will have to be automatic. When a dataset is too large for a particular algorithm to be applied, there are basically three ways to make learning feasible. The first one is trivial: instead of applying the scheme to the full dataset, use just a small subset of available data for training. Obviously, in this case information is easy to be lost and the loss is negligible in terms of correlation discovery between data. The second method consists of parallelization techniques. But the problem is to be able to derive a parallelized version of the learning algorithm. Sometimes it results feasible due to the intrinsic natural essence of the learning rule (such as genetic algorithms). However, parallelization is only a partial remedy because with a fixed number of available CPUs, the algorithm's asymptotic time complexity cannot be improved. Background knowledge (the third method) can make it possible to reduce the amount of data that needs to be processed by a learning rule. In some cases most of the attributes in a huge dataset might turn out to be irrelevant when background knowledge is taken into account. But in many exploration cases, especially related to data mining against data analysis problems, the background knowledge simply does not exists, or could infer a sort of wrong biased knowledge in the discovery process. In this scenario DAME (Data Mining & Exploration) project, starting from astrophysics requirements domain, has investigated the Massive Data Sets (MDS) exploration by producing a taxonomy of data mining applications (hereinafter called functionalities) and collected a set of machine learning algorithms (hereinafter called models). This association functionality-model represents what we defined as simply "use case", easily configurable by the user through specific tutorials. At low level, any experiment launched on the DAME framework, externally configurable through dynamical interactive web pages, is treated in a standard way, making completely transparent to the user the specific computing infrastructure used and specific data format given as input. As described in what follows, the result is a distributed data mining infrastructure, perfectly scalable in terms of data dimension and computing power requirements, originally tested and successfully validated on astrophysical science cases, able to perform supervised and unsupervised learning and revealing a multi-disciplinary data exploration capability. In practice, a specific software module of the Suite, called DRiver Management System (DRMS), that is a sub-system of the DRIVER (DR) component has been implemented to delegate at runtime the choice on which computing infrastructure should be launched the experiment. Currently the choice is between GRID or stand alone multi-thread platform, that could be replaced also by a CLOUD infrastructure, (but the DRMS is engineered in an easy expandable way, so it is also under further investigation the deployment of our Suite under a multi-core platform, based on GPU+CUDA computing technique). The mechanism is simple, being in terms of a threshold-based evaluation of the input dataset dimensions and the status of GRID job scheduler at execution startup time. This could reduce both execution time on a single experiment and the entire job execution scheduling.

## 2. MODERN E-SCIENCE REQUIREMENTS

E-science communities recently started to face the deluge of data produced by new generation of scientific instruments and by numerical simulations (widely used to model the physical processes and compare them with measured ones). Data is commonly organized in scientific repositories. Data providers have implemented Web access to their repositories, and http links between them. This *data network*, which consists of huge volumes of highly distributed, heterogeneous data, opened up many new research possibilities and greatly improved the efficiency of doing science. But it also posed new problems on the cross-correlation capabilities and mining techniques on these MDS to improve scientific results. The most important advance we expect is a dramatic need of the ease in using distributed e-Infrastructures for the e-science communities. We pursue a scenario where users sit down at their desks and, through a few mouse clicks, select and activate the most suitable *scientific gateway* for their specific applications or gain access to detailed documentation or tutorials. We call *scientific gateway* an e-Infrastructure which is able to offer remote access and navigation on distributed data repositories together with web services and applications able to explore, analyze and mine data. It doesn't require any software installation or execution on user local PC, it permits asynchronous connection and launch of jobs and embeds to the user any computing infrastructure configuration or management.

In this way the scientific communities will expand their use of the e-Infrastructure and benefit from a fundamental tool to undertake research, develop collaborations, and increase their scientific productivity and the quality of research outputs. Only if this scenario becomes reality, the barrier currently placed between the community of users and technology will disappear.

## 2.1 The case of Astrophysics

From the scientific point of view, the DAME project arises from the astrophysical domain, where the understanding of the universe beyond the Solar System is based on just a few information carriers: photons in several wavelengths, cosmic rays, neutrinos and gravitational waves. Each of these carriers has it peculiarities and weaknesses from the scientific point of view: they sample different energy ranges, endure different kinds and levels of interference during their cosmic journey (e.g. photons are absorbed while charged Cosmic Rays (CRs) are deflected by magnetic fields), sample different physical phenomena (e.g. thermal, non thermal and stimulated emission mechanisms), and require very different technologies for their detection. So far, the international community needs modern infrastructures for the exploitation of the ever increasing amount of data (of the order of PetaByte/year) produced by the new generation of telescopes and space borne instruments, as well as by numerical simulations of exploding complexity. Extending these requirements to other application fields, main goal of DAME project can be summarized in two items:

- The need of a "federation" of experimental data, by collecting them through several worldwide archives and by defining a series of standards for their formats and access protocols;
- The implementation of reliable computing instruments for data exploration, mining and knowledge extraction, user-friendly, scalable and as much as possible asynchronous;

These topics require powerful, computationally distributed and adaptive tools able to explore, extract and correlate knowledge from multivariate massive datasets in a multi-dimensional parameter space, **Fig. 1**.

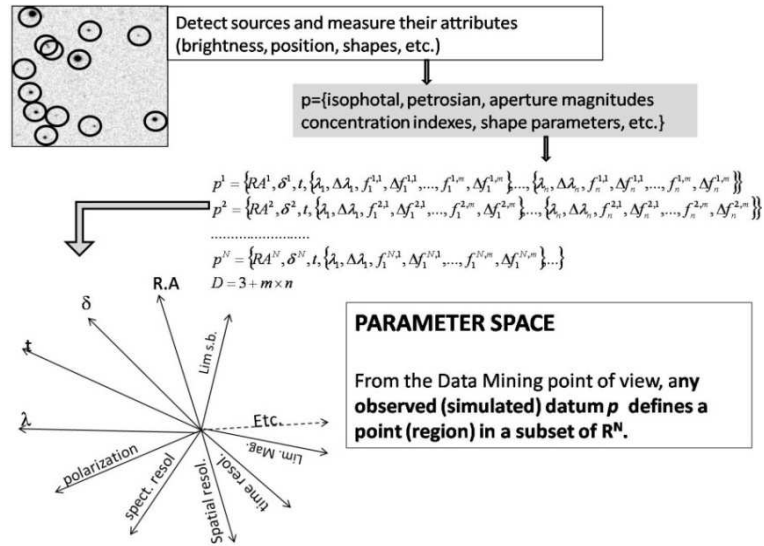

*Fig. 1 - The data multi-dimensional parameter space in Astrophysics problems*

The latter results as a typical data mining requirement, dealing with many scientific, social and technological environments. Concerning the specific astrophysical aspects, the problem, in fact, can be analytically expressed as follows. Any observed (or simulated) datum defines a point (region) in a subset of $R^N$, such as: R.A., DEC, time, wavelength, experimental setup (i.e. spatial and/or spectral resolution, limiting magnitude, brightness, etc.), fluxes, polarization, spectral response of the instrument and PSF;

Every time a new technology enlarges the parameter space or allows a better sampling of it, new discoveries are bound to take place. So far, the scientific exploitation of a multi-band (D bands), multi-epoch (K epochs) universe implies to search for patterns and trends among N points in a DxK dimensional parameter space, where $N > 10^9$, $D >> 100$, $K > 10$. The problem also requires a multi-disciplinary approach, covering aspects belonging to Astronomy, Physics, Biology, Information Technology, Artificial Intelligence, Engineering and Statistics environments. In the last decade, the Astronomy & Astrophysics communities participated in a number of initiatives related to the use and development of e-Infrastructures for science and research (e.g. EGEE [7], EuroVO [8], grid.IT [9]), giving astronomers the possibility to develop well established and successful VRC (Virtual Research Communities). A support cluster dedicated to A&A has been set up and funded in the framework of the EGEE-III project. Surveys of the requirements of the A&A community concerning data management, job management, and distributed tools and more general services has been done in the framework of EGEE-II and EGEE-III projects. Requirements focus on the need to integrate astronomical databases in a computing grid and create proper science gateways to underpin the use of the infrastructure and to bridge heterogeneous e-Infrastructures (i.e. EGEE and EuroVO). Moreover, astronomers envisage some advanced functionalities in the use of new computing architectures (such as shared memory systems or GPU computing) and therefore the ability to *gridify* applications that require to run many independent tasks of parallel jobs. This dynamic process demonstrates the effectiveness of the two basic requirements mentioned above.

As for data, the concept of "distributed archives" is already familiar to the average astrophysicist. The leap forward in this case is to be able to organize the data repositories to allow efficient, transparent and uniform access: these are the basic goals of the VO or VObs (Virtual Observatory). In more than a sense, the VO is an extension of the classical Computational Grid; it fits perfectly the Data Grid concept, being based on storage and processing systems, and metadata and communications management services. The VO is a paradigm to use multiple archives of astronomical data in an interoperating, integrated and logically centralized way, so to be able to "observe a virtual sky"' by position, wavelength and time. Not only data actually observed are included in this concept: theoretical and diagnostic can be included as well. VO represents a new type of a scientific organization for the era of information abundance:

- It is inherently *distributed*, and web-centric;
- It is fundamentally based on a *rapidly developing technology*;
- *It transcends the traditional boundaries* between different wavelength regimes, agency domains;
- It has an *unusually broad range of constituents* and interfaces;
- It is inherently *multidisciplinary*;

The International VO (cf. the IVO Alliance or IVOA, [10]) has opened a new frontier to astronomy. In fact, by making available at the click of a mouse an unprecedented wealth of data and by implementing common standards and procedures, the VObs allow a new generation of scientists to tackle complex problems which were almost unthinkable only a decade ago [1]. Astronomers may now access a "virtual" parameter space of increasing complexity (hundreds or thousands features measured per object) and size (billions of objects). However, the link between data mining applications and the VObs is currently defined only partially. As a matter of fact, IVOA has concentrated its standardization efforts up to now mainly on data, and the definition of mechanisms to access general purpose interoperable tools for "server side" massive data sets manipulation is still a matter of discussion within IVOA. Behind this consideration there is the crucial requirement to harmonize all recent efforts spent in the fields of VObs, GRID and HPC computing, and data mining.

## 3. DATA MINING AND THE FOURTH PARADIGM OF SCIENCE

X-informatics (such as Bio-informatics, Geo-informatics and Astro-informatics), is growingly being recognized as the fourth leg of scientific research after experiment, theory and simulations [2]. It arises from the pressing need to acquire the multi-disciplinary expertise which is needed to deal with the ongoing burst of data complexity and to perform data mining and exploration on MDS. The crucial role played by such tasks in astrophysics research has been recently certified by the constitution, within the IVOA, of an Interest Group on Knowledge Discovery in Data Bases (KDD-IG) which is seen as the main interface between the IVOA technical infrastructure and the VO enabled science. In this context the DAME project intends:

- To provide the VO with an extensible, integrated environment for data mining and exploration;
- Support of the VO standards and formats, especially for application interop (SAMP);
- To abstract the application deployment and execution, so to provide the VO with a general purpose computing platform taking advantage of the modern technologies (e.g. Grid, Cloud, etc...).

By following the fourth paradigm of science, it is now emerging world-wide (cf. the US – AVO community & the recent meeting on Astro-informatics at the 215th AAS) the need for all components (both hardware and software) of the Astro-informatics infrastructure to be integrated or, at least, made fully interoperable. In other words, the various infrastructure components (data, computational resources and paradigms, software environments, applications) should interact seamlessly exchanging information, and be based on a strong underlying network component.

# 4. THE DAME APPROACH TO DISTRIBUTED DATA MINING

The DAME project aims at creating a distributed e-infrastructure to guarantee integrated and asynchronous access to data collected by very different experiments and scientific communities in order to correlate them and improve their scientific usability. The project consists of a data mining framework with powerful software instruments capable to work on MDS in a distributed computing environment. The VObs have defined a set of standards to allow interoperability among different archives and databases in the astrophysics domain, and keeps them updated through the activity of dedicated working groups. So far, most of the implementation effort for the VO has concerned the storage, standardization and interoperability of the data together with the computational infrastructures. Our project extends this fundamental target by integrating it in an infrastructure, joining service-oriented software and resource-oriented hardware paradigms, including the implementation of advanced tools for KDD purposes. The DAME design takes also into account the fact that the average scientists cannot and/or does not want to become an expert also in Computer Science or in the fields of algorithms and ICT. In most cases the r.m.s. scientist (our end user) already possesses his own algorithms for data processing and analysis and has implemented private routines/pipelines to solve specific problems. These tools, however, often are not scalable to distributed computing environments or are too difficult to be migrated on a GRID infrastructure. DAME also aims at providing a user friendly scientific gateway to easy the access, exploration, processing and understanding of the massive data sets federated under standards according VObs rules. We wish to emphasize that standardization needs to be extended to data analysis and mining methods and to algorithms development. The natural computing environment for such MDS processing is a distributed infrastructure (GRID/CLOUD), but again, we need to define standards in the development of higher level interfaces, in order to:

- isolate end user from technical details of VO and GRID/CLOUD use and configuration;
- make it easier to combine existing services and resources into experiments;

Data Mining is usually conceived as an application (deterministic/stochastic algorithm) to extract unknown information from noisy data. This is basically true but in some way it is too much reductive with respect to the wide range covered by mining concept domains. More precisely, in DAME, data mining is intended as techniques of exploration on data, based on the combination between parameter space filtering, machine learning, soft computing techniques associated to a functional domain. The functional domain term arises from the conceptual taxonomy of research modes applicable on data. Dimensional reduction, classification, regression, prediction, clustering, filtering are example of functionalities belonging to the data mining conceptual domain, in which the various methods (models and algorithms) can be applied to explore data under a particular aspect, connected to the associated functionality scope.

## 4.1 Design Architecture

DAME is based on five main components: Front End (FE), Framework (FW), Registry and Data Base (REDB), Driver (DR) and Data Mining Models (DMM).

The FW is the core of the Suite. It handles all communication flow from/to FE (i.e. the end user) and the rest of the components, in order to register the user, to show user working session information, to configure and execute all user experiments, to report output and status/log information about the applications running or already finished. One of the most critical factors of the FW component is the interaction of a newly created experiment with the GRID environment. The FW needs to create and configure the plug-in (hereinafter called DMPlugin) associated to the experiment. After the DMPlugin is configured the DR component needs to run the experiment by calling the run method of the plug-in. When executed on the GRID, the process needs to migrate on a Worker Node (WN). To implement this migration we've chosen to serialize the DMPlugin in a file. Serialization is a process of converting an object into a sequence of bits so that it can be stored on a storage medium. Our tests on the GRID environment indicates that this solution works fine and that the jdl file needed to manage the whole process is very simple.

The component FE includes the main GUI (Graphical User Interface) of the Suite and it is based on dynamical WEB pages, rendered by the Google Web Toolkit (GWT), able to interface the end users with the applications, models and facilities to launch scientific experiments. The interface foresees an authentication procedure which redirects the user to a personal session environment, collecting uploaded data, check experiment status and driven procedures to configure and execute new scientific experiments, using all available data mining algorithms and tools. From the engineering point of view, the FE is organized by means of a bidirectional information exchange, through XML files, with the component FW, suite engine component, as showed in **Fig. 2**.

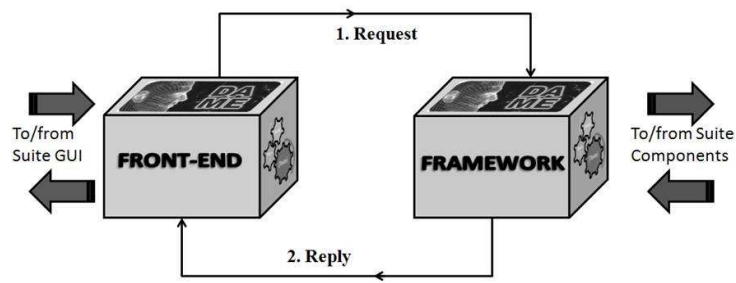

*Fig. 2 –communication interface schema between FE and FW*

The component DR is the package responsible of the physical implementation of the HW resources handled by other components at a virtual level. It permits the abstraction of the real platform (HW environment and related operative system calls) to the rest of Suite software components, including also I/O interface (file loading/storing), user data intermediate formatting and conversions (ASCII, CSV, FITS, VO-TABLE), job scheduler, memory management and process redirection (**Fig. 3**).

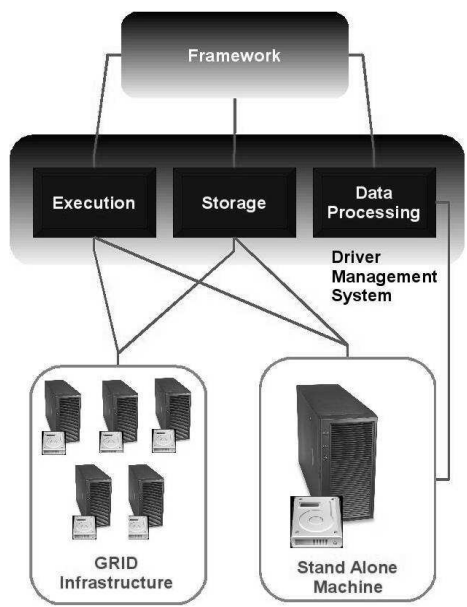

*Fig.3 - The DRIVER component as interface with computing infrastructure*

More in detail, a specific sub-system of the DR component, called DRiver Management System (DRMS), has been implemented to delegate at runtime the choice of the computing infrastructure should be selected to launch the experiment.

The component REDB is the base of knowledge repository for the Suite. It is a registry in the sense that contains all information related to user registration and accounts, his working sessions and related experiments.

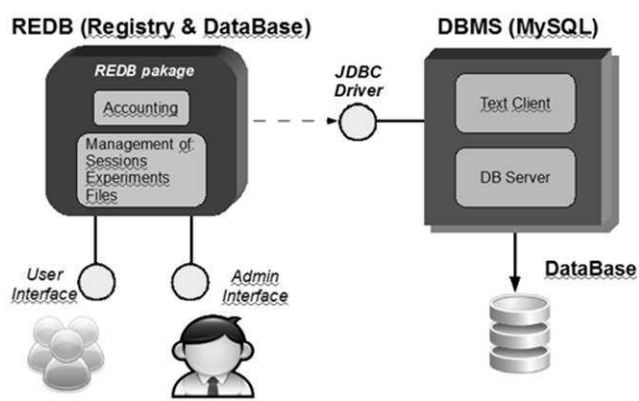

*Fig. 4 – REDB architecture*

It is also a Database containing information about experiment input/output data and all temporary/final data coming from user jobs and applications, (**Fig. 4**).

The component DMM is the package implementing all data processing models and algorithms available in the Suite. They are referred to supervised/unsupervised models, coming from Soft Computing, Self-adaptive, Statistical and Deterministic computing environments. It is structured by means of a package of libraries (Java API) referred to the following items:

- Data mining models libraries (Multi Layer Perceptron, Support Vector Machine, Genetic Algorithms, Self Organizing Maps, etc…);
- Visualization tools;

- Statistical tools;
- List of functionalities (Classification, Regression, Clustering, etc…);
- Custom libraries required by the user;

The following scheme shows the component diagram of the entire suite (**Fig. 5**).

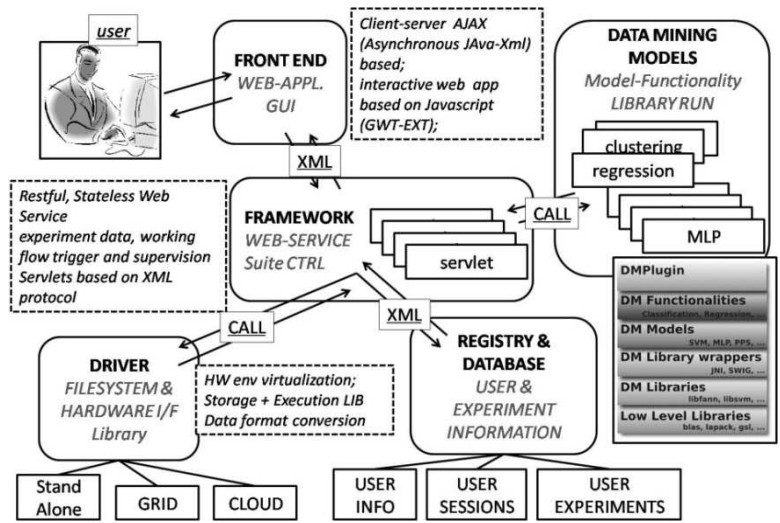

*Fig. 5 - DAME Functional Infrastructure*

## 4.2 Distributed Environment

As underlined in the previous sections processing of huge quantities of data is a typical requirement of e-science communities. The amount of computations needed to process the data is impressive, but often "embarrassingly parallel" since based on local operators, with a coarse-grained level of parallelism. In such cases, the "memory footprint" of the applications allows to subdivide data in chunks, so as to fit the RAM available on the individual CPUs and to have each CPU to perform a single processing unit. In most cases "distributed supercomputers", i.e. a local cluster of PCs such as a Beowulf machine, or a set of HPC computers distributed over the network, can be an effective solution to the problem. In this case, the GRID paradigm can be considered to be an important step forward in the provision of the computing power needed to tackle the new challenges.

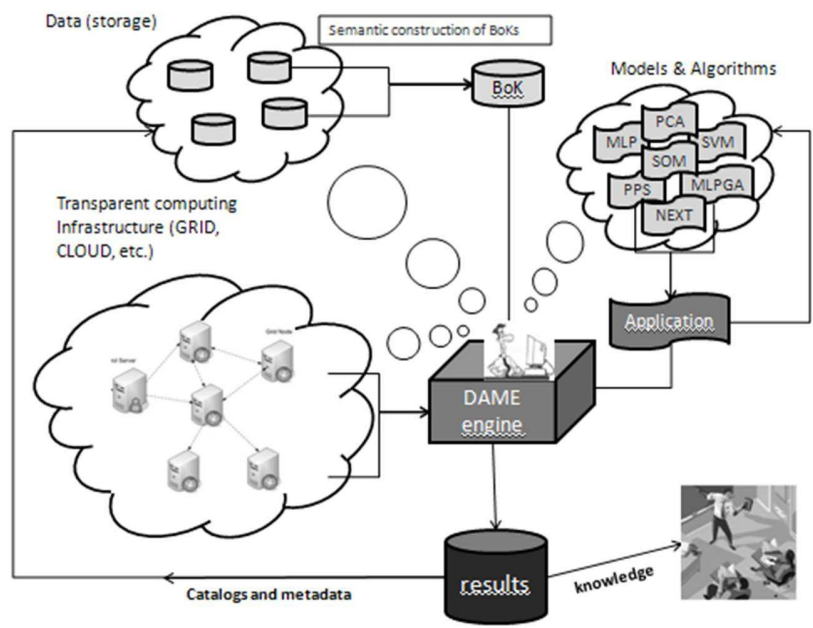

*Fig. 6 - The concept of distributed infrastructure in DAME*

The main concept of distributed data mining applications embedded in the DAME package Suite is based on three issues (**Fig. 6**):

- Virtual organization of data: this is the extension of already remarked basic feature of VObs;
- Hardware resource-oriented: this is obtained by using computing infrastructures, like GRID, whose solutions enable parallel processing of tasks, using idle capacity. The paradigm in this case is to obtain large numbers of work requests running for short periods of time;
- Software service-oriented: this is the base of typical CLOUD computing paradigm. The data mining applications implemented runs on top of virtual machines, seen at the user level as services (specifically web services), standardized in terms of data management and working flow;

Our scientific community needs not only "traditional" computations but also the use of complex data operations that require on-line access to databases mainly mediated through a set of domain specific web services (e.g. VObs), and the use of HPC resources to run *in-silico* (numerical) experiments. The DAME Suite is deployed on an multi-environment platform including both CLOUD and GRID solutions (**Fig. 7**).

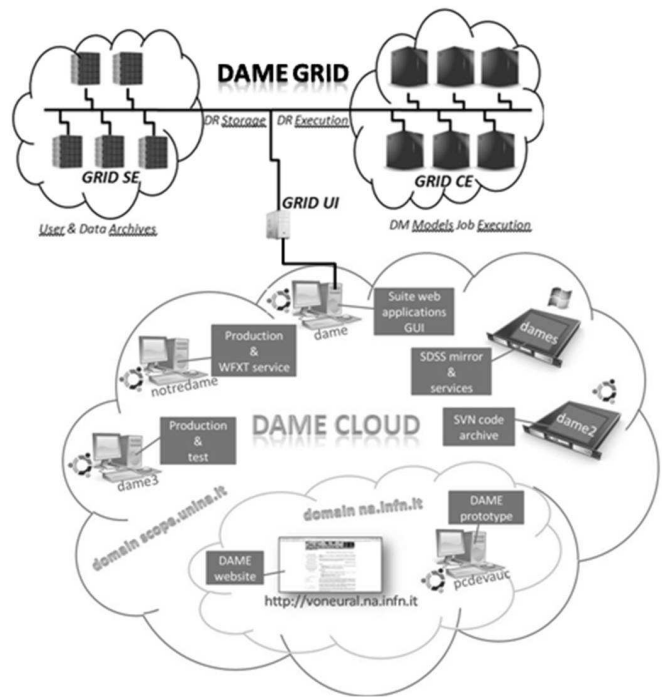

*Fig. 7 - The DAME Suite deployed on the GRID architecture*

In particular, concerning the GRID side, the Suite exploits the S.Co.P.E. GRID infrastructure. The S.Co.P.E. project [3], aimed at the construction and activation of a Data Center which is now perfectly integrated in the national and international GRID initiatives, hosts 300 eight-core blade servers and 220 Terabyte of storage. The acronym stands for "Cooperative System for Multidisciplinary Scientific Computations", that is a collaborative system for scientific applications in many areas of research. For its generality the DAME Suite is used also for applications outside of astronomy (such as, chemistry, bioinformatics and social sciences).

## 4.3 Soft Computing Applications

The KDD scheme adopted in the DAME package is based on Soft Computing methods, belonging to the typical dichotomy (supervised/unsupervised) of machine learning methods. First type makes use of prior knowledge to group samples into different classes. In the second type, instead, null or very little a priori knowledge is required and the patterns are classified using only their statistical properties and some similarity measure which can be quantified through a mathematical clustering objective function, based on a properly selected distance measure. In the first release, the DMM implements the models as listed in the following table.

| MODEL | CATEGORY | FUNCTIONALITY |
|---|---|---|
| MLP + Back Propagation learning rule, [11] | Supervised | Classification, Regression |
| MLP with GA learning rule, [11], [16] | Supervised | Classification, Regression |
| SVM, [14] | Supervised | Classification, Regression |
| SOM, [15] | Unsupervised | Clustering |
| Principal Probabilistic Surfaces (PPS), [13] | Unsupervised | Dimensional reduction, pre-clustering |
| Negative Entropy Clustering (NEC), [12] | Unsupervised | Clustering |

Depending on the specific experiment, the use of any of the models listed above can be executed in a more or less degree of parallelization. All the models require some parameters that cannot be defined a priori, causing the necessity of iterated experiment sessions in order to find the best tuning. Then not all the models can be developed under the Message Passing Interface (MPI) paradigm. But the possibility to execute more jobs at once (specific GRID case) intrinsically exploits the multi-processor architecture.

## 5. FIRST SCIENTIFIC AND TECHNOLOGICAL RESULTS

During the design and development phases of the project, some prototypes have been implemented in order to verify the project issues and to validate the selected data mining models from the scientific point of view. In this context, a Java-based plugin wizard for custom experiment (DMPlugin) setup has been designed to extend DAME Suite features with user own algorithms to be applied to scientific cases by encapsulating them inside the Suite (**Fig. 8**).

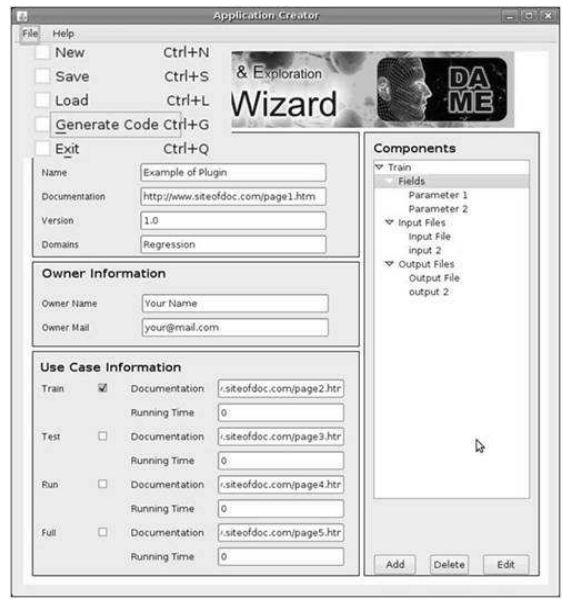

*Fig. 8 - The DMPlugin Java application to extend functionalities of DAME Suite*

This facility, [17], extends the canonical use of the Suite: a simple user can upload and build his datasets, configure the data mining models available, execute different experiments in service mode, load graphical views of partial/final results.

Moreover, a prototype of the framework Suite has been developed (**Fig. 9**). The prototype is a Web Application implementing minimal DAME features and requirements, developed in parallel with the project advancement in order to perform a scientific validation of models and algorithms foreseen in the main Suite and to verify all basic project features designed related to the scientific pipeline workflow. It has been implemented as a Python web application and is publicly accessible, [18].

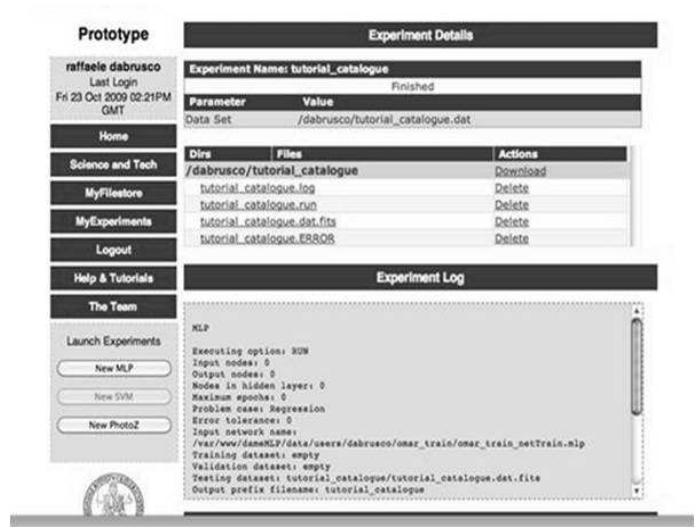

*Fig. 9 - The DAME Prototype page example*

The prototype implements the basic user interface functionalities: a virtual file store for each registered user is physically allocated on the machine that serves the web application. Users can upload their files, delete them, visualize them: the system tries to recognize the file type and shows images or text contextually. Any astronomical data analysis and/or data mining experiment to be executed on the prototype, can be organized as a data processing pipeline, in which the use of the prototype needs to be integrated with pre and post processing tools, available between Virtual Observatory Web Services. The prototype has been tested on three different science cases which make use of MDS:

- **Photometric redshifts for the SDSS galaxies, [19]:** It makes use of a nested chain of MLP (Multi Layer Perceptron) and allowed to derive the photometric redshifts for ca. 30 million SDSS galaxies with an accuracy of 0.02 in redshift. This result which has appeared in the Astrophysical Journal [4], was also crucial for a further analysis of low multiplicity groups of galaxies (Shakhbazian) in the SDSS sample;

- **Search for candidate quasars in the SDSS:** The work was performed using the PPS (Probabilistic Principal Surfaces) module applied to the SDSS and SDSS+UKIDS data. It consisted in the search for candidate quasars in absence of a priori constrains and in a high dimensionality photometric parameter space, [5];

- **AGN classification in the SDSS [23]**: Using the GRID-S.Co.P.E. to execute 110 jobs on 110 WN, the SVM model is employed to produce a classification of different types of AGN using the photometric data from the SDSS and the base of knowledge provided by the SDSS spectroscopic subsamples. A paper on the results is in preparation.

Concerning the results achieved and remarking what already mentioned in par. 5.3, using the hybrid architecture, it is possible to execute simultaneous experiments that gathered all together, bring the best results. Even if the single job is not parallelized, we obtain a running time improvement by reaching the limit value of the Amdahl's law (N):

$$\frac{1}{(1-P)+\frac{P}{N}},$$

where if P is the proportion of a program that can be made parallel (i.e. benefit from parallelization), and (1 – P) is the proportion that cannot be parallelized (remains serial), then the resulting maximum speed up that can be achieved by using N processors is obtained by the law expressed above.

For example, in the case of AGN Classification experiment (cited above), each of the 110 jobs runs for about a week on a single processor. By exploiting the GRID, the experiment running time can be reduced to about one week instead of more than 2 years (110 weeks).

## 6. CONCLUSION

Generally speaking, applications for KDD will come not from computer programs, nor from machine learning experts, nor from the data itself, but from people and communities who work with the data and the problems from which it arises. That is why we have designed and provided the DAME infrastructure, to empower those who are not machine learning experts to apply these techniques to the problems that arise in daily working life. DAME project comes out as an astrophysical data exploration and mining tool, originating from the very simple consideration that, with data obtained by the new generation of instruments, we have reached the physical limit of observations (single photon counting) at almost all wavelengths. If extended to other scientific or applied research disciplines, the opportunity to gain new insights on the knowledge will depend mainly on the capability to recognize patterns or trends in the parameter space, which are not limited to the 3-D human visualization, from very large datasets. In this sense DAME approach can be easily and widely applied to other scientific, social, industrial and technological scenarios. Our project has recently passed the R&D phase, de facto entering in the implementation commissioning step and by performing in parallel the scientific testing with first infrastructure prototype, accessible, after a simple authentication procedure, through the official project website address, [20]. First scientific test results confirm the goodness of the theoretical approach and technological strategy.

## ACKNOWLEDGEMENT

DAME is funded by the Italian Ministry of Foreign Affairs as well as by the European project VOTECH (Virtual Observatory Technological Infrastructures, [21]), and by the Italian PON-S.Co.P.E., [22]. Official partners of the project are:

- Department of Physics - University Federico II of Napoli, Italy
- INAF National Institute of Astrophysics – Astronomical Observatory of Capodimonte , Italy
- California Institute of Technology, Pasadena, USA